\documentclass[10pt,twoside,onecolumn,a4paper]{article}

\usepackage{graphicx}  
\usepackage{dcolumn}   
\usepackage{bm}        
\usepackage{amssymb}   

\begin{document}

\title{Optical frequency transfer via a 660~km underground fiber link using a remote Brillouin amplifier}
\author{S. M. F. Raupach, A. Koczwara, and G. Grosche\\Physikalisch-Technische Bundesanstalt (PTB), Bundesallee 100,\\38116 Braunschweig, Germany\\ \textit{\small{e-mail: sebastian.raupach@ptb.de}}}
 
\maketitle

\begin{abstract}
In long-haul optical continuous-wave frequency transfer via fiber, remote bidirectional Er$^+$-doped fiber amplifiers are commonly used to mitigate signal attenuation. We demonstrate for the first time the ultrastable transfer of an optical frequency using a remote fiber Brillouin amplifier, placed in a server room along the link. Using it as the only means of remote amplification, on a 660~km loop of installed underground fiber we bridge distances of 250~km and 160~km between amplifications. Over several days of uninterrupted measurement we find an instability of the frequency transfer (Allan deviation of $\Lambda$-weighted data with 1 s gate time) of around $1\times10^{-19}$ and less for averaging times longer than 3000~s. The modified Allan deviation reaches $3\times10^{-19}$ at an averaging time of 100~s, corresponding to the current noise floor at this averaging time. For averaging times longer than 1000~s the modified Allan deviation is in the $10^{-20}$ range. A conservative value of the overall accuracy is $1\times10^{-19}$.
\end{abstract}

\section{Introduction}
The low-noise transfer of ultrastable continuous-wave optical frequencies is a highly active field of research \cite{predehl2012,lopez2012frequency,droste2013,calonico2014}. It is a prerequisite for comparing the emerging new generation of atomic clocks based on optical transitions \cite{chou2010,nicholson2012,hinkley2013,falke2013,bloom2014}. These show low instabilities and accuracies in the $10^{-17}$ range and better. Traditional means of satellite frequency transfer using microwave carriers \cite{bauch2006},  would in theory require decades of averaging time to regain the transmitted signal from the noise of the transmission at these levels of stability, if possible at all. Further applications of low-noise optical frequency transfer include remote spectroscopy and laser characterization \cite{pape2010}, or remote synchronisation via ultrastable chirped frequency transfer \cite{raupach2014}.
\\Long distance transfer of optical frequencies typically relies on transmitting an optical infrared frequency via underground, standard single-mode silica fiber telecommunication networks \cite{predehl2012,lopez2012frequency,droste2013,calonico2014,williams2008,calosso2014}. It has been demonstrated over fiber lengths close to 2000 km \cite{droste2013}. One challenge is the mitigation and suppression of the noise imprinted onto the signal’s phase by acoustic/mechanical and thermal perturbations to the fibers \cite{williams2008}. This can be done in real-time using interferometric stabilization \cite{predehl2012,lopez2012frequency,droste2013,calonico2014,williams2008} or via post-processing of the recorded data, e.g. in a two-way transfer scheme \cite{calosso2014}. 
\\Another challenge is the attenuation of the signal when transmitted over fiber lengths in excess of about 100 km. In the telecommunication wavelength window around 1550 nm, standard silica fiber exhibits an attenuation around 0.2 dB/km. Additional losses occur e.g. at splices and connectors, therefore the effective attenuation can be as high as 28\ldots30 dB per 100 km \cite{lopez2012frequency,calonico2014}.
\\In the context of ultrastable transfer of optical frequencies, different approaches to amplify or regenerate the signal are investigated or employed. These include the use of Er$^+$-doped fiber amplifiers (EDFA) as the most common one \cite{predehl2012,lopez2012frequency,calonico2014}. Other approaches investigate Raman amplification \cite{clivati2013} and Brillouin amplification \cite{terra2010,raupach2013}, where the transmission fiber itself is used as a gain medium, or phase-lock a repeater laser to the incoming signal \cite{lopez2012frequency}. Injection locking a slave laser to the incoming signal might also be a possibility \cite{wu2013}. In France and Germany, also mixed-amplifier links are operated, where EDFA and in-lab Brillouin pumps \cite{predehl2012} or EDFA and signal repeaters \cite{lopez2012frequency} are employed to mitigate excess losses of the transfer.
\\\\Both the interferometric as well as the two-way approach require bidirectional, reciprocal transmission. Therefore, contrary to telecommunication applications, in ultrastable frequency transfer the amplifiers cannot include optical isolators. When using broad-band amplifiers such as EDFA, this restricts the gain to below 20 dB \cite{lopez2012frequency} to avoid spontaneous lasing due to Rayleigh back-scattering or back-reflections. Thus the amplification does not fully match the attenuation mentioned above. Consequently, over long links excess losses will accumulate unless the fiber path between two adjacent EDFA is well below 100 km.  Also, the occurrence of several signal failures or cycle slips in the detected signal per hour observed in optical frequency transfer \cite{predehl2012,droste2013,calonico2014} may be related to the intricacies of cascaded bidirectional operation of EDFA.
\\Fiber Brillouin amplification on the other hand works only for counterpropagating pump and signal and has a narrow gain bandwidth of order 10 MHz \cite{subias2009,rohde2013}. The latter is challenging in terms of frequency accuracy of the pump light relative to the signal light, which is one of the reasons why up to now it was restricted to laboratory applications only. On the other hand, it allows to individually amplify either the outgoing or the frequency-shifted incoming light. This allows exploiting small-signal gains in excess of 40 dB.
\\At Physikalisch-Technische Bundesanstalt (PTB), a field-able fiber Brillouin amplifier suited for remote operation is being developed \cite{raupach2013} with support from the European Metrology Programme (EMRP). It is now operated on a 660 km loop of installed underground fiber.
\\In the first section  we will describe the experimental setup of the loop measurement including the remote fiber Brillouin amplifier. We will then present measurements of the phase noise and instability observed in the transfer of an optical frequency, before finally discussing the results.

\section{Loop setup including one remote fiber Brillouin amplifier}
\begin{figure*}[htbp]
\centering\includegraphics[width=15cm]{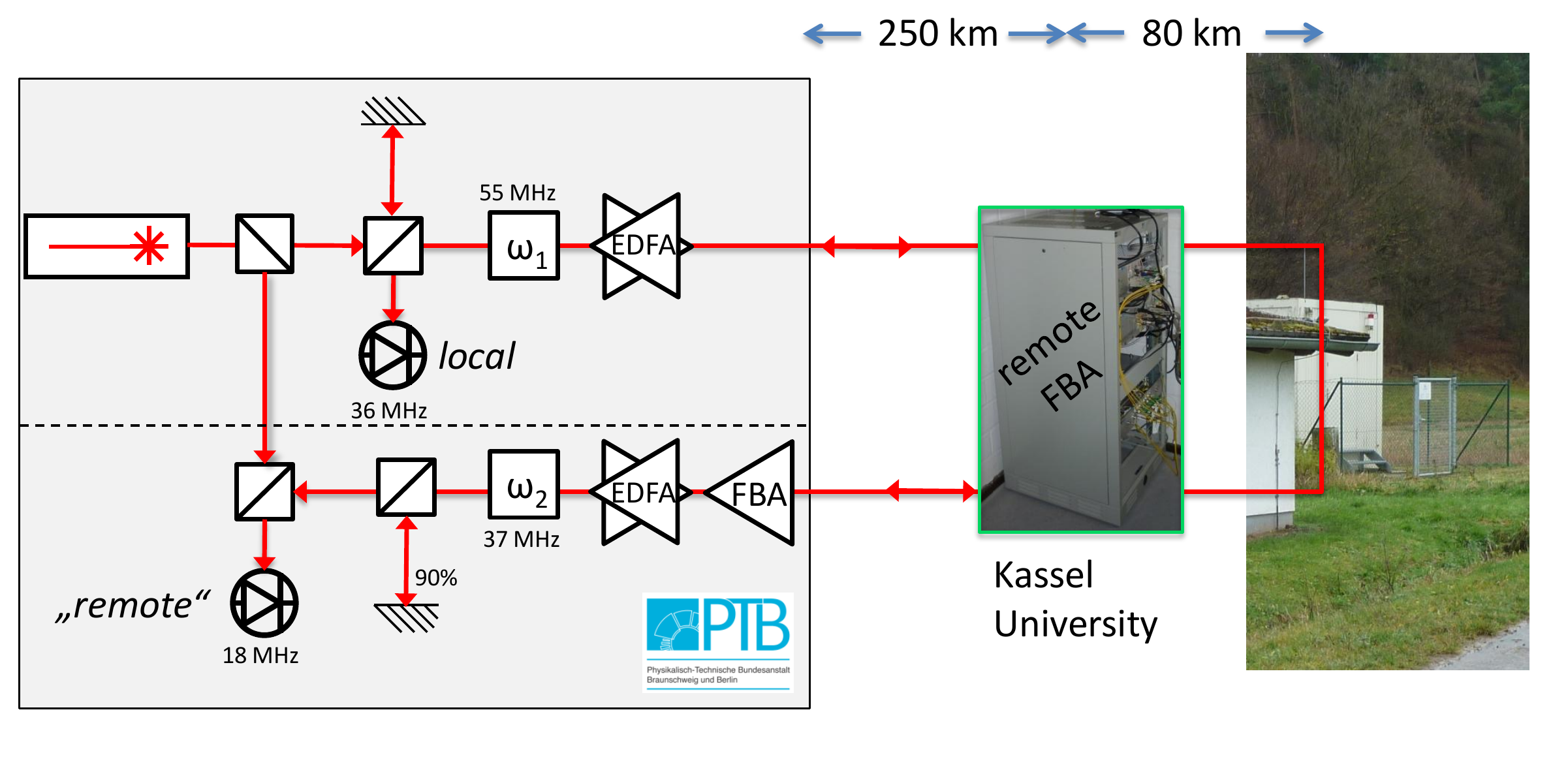}
\caption{Schematic sketch of the 660 km loop setup PTB-PTB, including one remote fiber Brillouin amplifier (FBA); EDFA: Erbium-doped fiber amplifier; FBA: fiber Brillouin amplifier; $\omega_{1,2}$: acousto-optical modulator.} 
\label{fig:loop}
\end{figure*}

\begin{figure*}[htbp]
\centering\includegraphics[width=8cm]{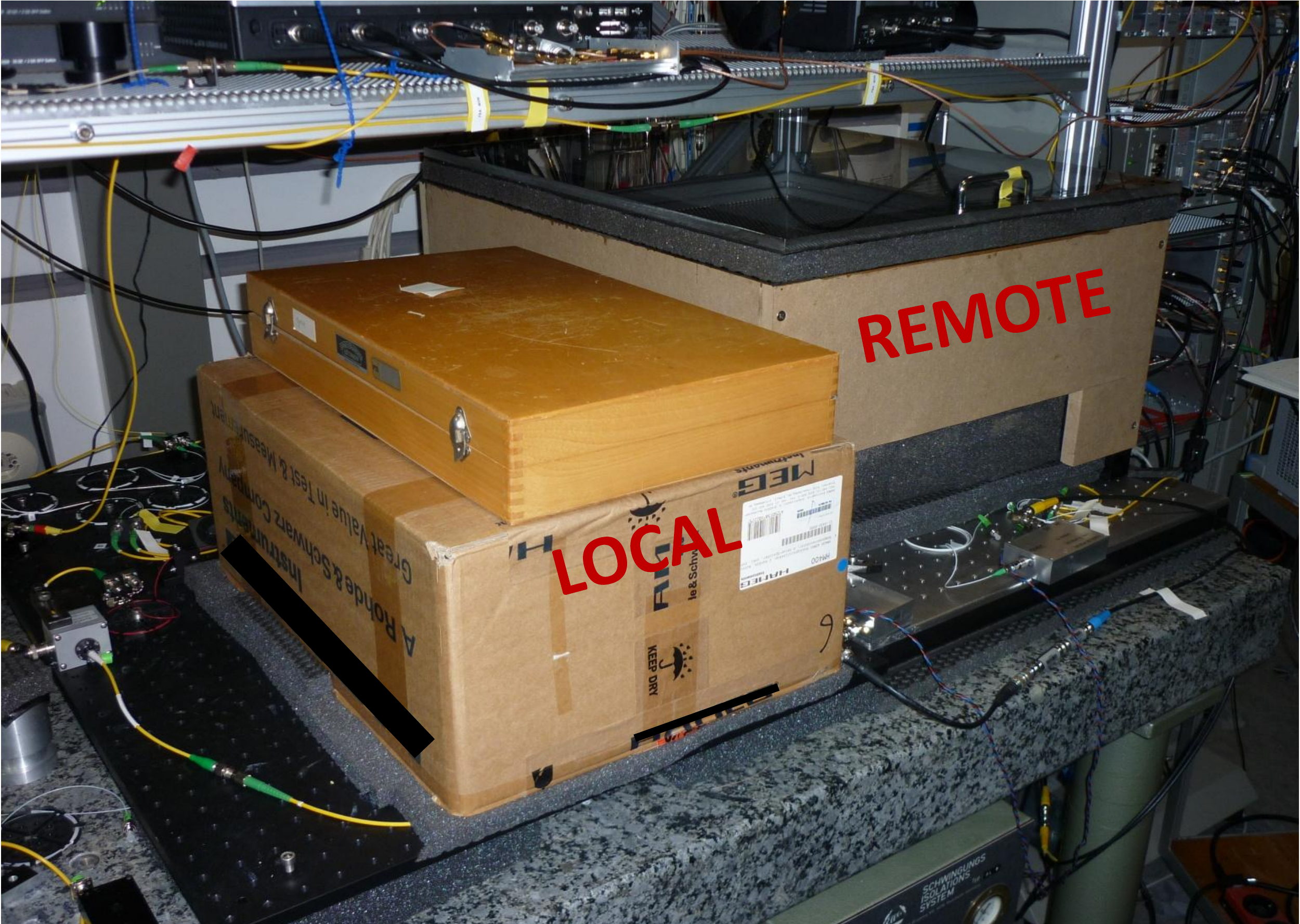}
\caption{The interferometer used for stabilizing the loop using the inloop beat and the optics for generating the ``remote'' beat are placed in adjacent but separate, thermally insulated housings.} 
\label{fig:boxes}
\end{figure*}

The experimental setup is shown schematically in figure \ref{fig:loop}. The transfer laser is a continuous-wave (cw) fiber laser (Koheras Adjustik), phase-locked to an ULE-cavity stabilized master laser having Sub-Hz linewidth. The transfer laser has a frequency of 194.4 THz, corresponding to ITU-channel 44. Its light is split into different paths for e.g. locking the local Brillouin pump laser to the outgoing signal light. The transmitted signal's phase is stabilized actively using interferometric stabilization \cite{predehl2012,lopez2012frequency,calonico2014,williams2008}: The setup forms a Michelson interferometer, where the link itself forms the object path of the interferometer.  The beat signal of the roundtrip light in  the radio frequency (rf) range is detected and compared to the signal of a local rf  oscillator derived from a hydrogen maser.  A phase-locked-loop suppresses the detected phase deviations by acting on the acousto-optical modulator ``$\omega_1$'', thus stabilizing the mean phase and frequency at the ``remote'' end. An acousto-optic modulator ``$\omega_2$''at the remote end applies a fixed frequency shift to the signal to distinguish it from light merely back-scattered along the link. At the same time, it shifts the signal's frequency into the gaincurve of the Brillouin pumps for the returning light.
\\The link itself consists of a pair of standard single-mode telecom fiber, forming part of a future fiber connection between Braunschweig and Paris. To allow convenient characterisation of the transfer, the two fibers are patched together in the federal state of Hassia between the cities of Kassel and Marburg, such that they form a loop starting and ending in the same laboratory at PTB. The length of this fiber loop is around 660 km.
While the local and the “remote” setup are collocated in the same laboratory with active climate control, care has been taken to put the local interferometer and the “remote” optics for generating the beatnote into separate, thermally insulated (no active temperature control) housings onto separate aluminum supports, see fig. \ref{fig:boxes}. This approximates a situation where the ends are not collocated, as in the operation of a point-to-point link.
\\\\At Kassel University, at a distance (fiber length) of around 250 km from PTB a remote Brillouin amplifier is installed and operated in a server room of the computing centre; the geographical distance is around 130 km. The pump light is phase-locked to the incoming signal at an offset around 11 GHz, corresponding to the Brillouin shift given by the silica fiber. The remote fiber Brillouin amplifier performs a total of four Brillouin amplifications of the outgoing and returning signal in the fiber loop. At the ``remote'' end (at the lab in PTB), a fifth, in-lab Brillouin pump amplifies the ``remote'' signal. Here the pump is offset-locked locally to the signal laser. At the ``remote'' end the transferred light is combined with that of the signal laser to obtain the “remote” beat signal for out-of-loop characterisation of the frequency transfer. Also, the setup comprises in-lab bidirectional EDFAs as signal boosters at the local and “remote” end, because of considerable on-campus losses and excess losses on the fiber PTB-Kassel. The beat frequencies are each tracked by two tracking oscillators in parallel and are measured by dead-time free K\&K counters. To characterize the frequency transfer, we measure the phase noise of the transferred light relative to the outgoing signal light, as well as its frequency instability and the deviation from the expected frequency offset given by the frequencies of the acousto-optical modulators.
\\\\The remote Brillouin amplifier setup consists of an externally amplified DFB laser with a free-running linewidth of several kilohertz as a pump source. Its output is split into four branches who serve two directions into each of the two fibers.  Two of the branches are shifted in frequency to accomodate the signal's frequency shift on the returning path. The pump power injected into the fiber is around 10 mW. The four branches include electrically controlled polarization controllers, for automated or remote optimization of the pump polarizations, taking advantage of an optical communication channel around 30 nm off the frequency-signal wavelength. Optimization is done by maximizing the amplified signal's intensity. On some days we observe fast fluctuations of the polarization, which we attribute to maintenance work along the link; they may cause a loss of lock of the remote amplifier and may warrant further development of the polarization optimization.
\\Using a beat note around 11 GHz between the pump light and the signal coming in from the local end of the link, the pump is phase locked to the signal. Locking can either be performed remotely, or automatically using a locking algorithm. For automated locking, the pump frequency is scanned to find the maximum of the amplified signal intensity followed by increasing the amplifier’s control loop gain for achieving the phase lock.
\\The fibers at the Kassel location were chosen such that the Brillouin shifts in all directions are the same to within less than 10~MHz. A central control unit allows for remote access as well as for largely autonomous operation of the amplifier setup.  The autonomous operation includes the active mitigation of temperature changes by acting on the DFB pump laser using control software developed in-house.
 
\section{Instability and phase noise of the transferred optical frequency}
\begin{figure*}[htbp]
\centering\includegraphics[width=8cm]{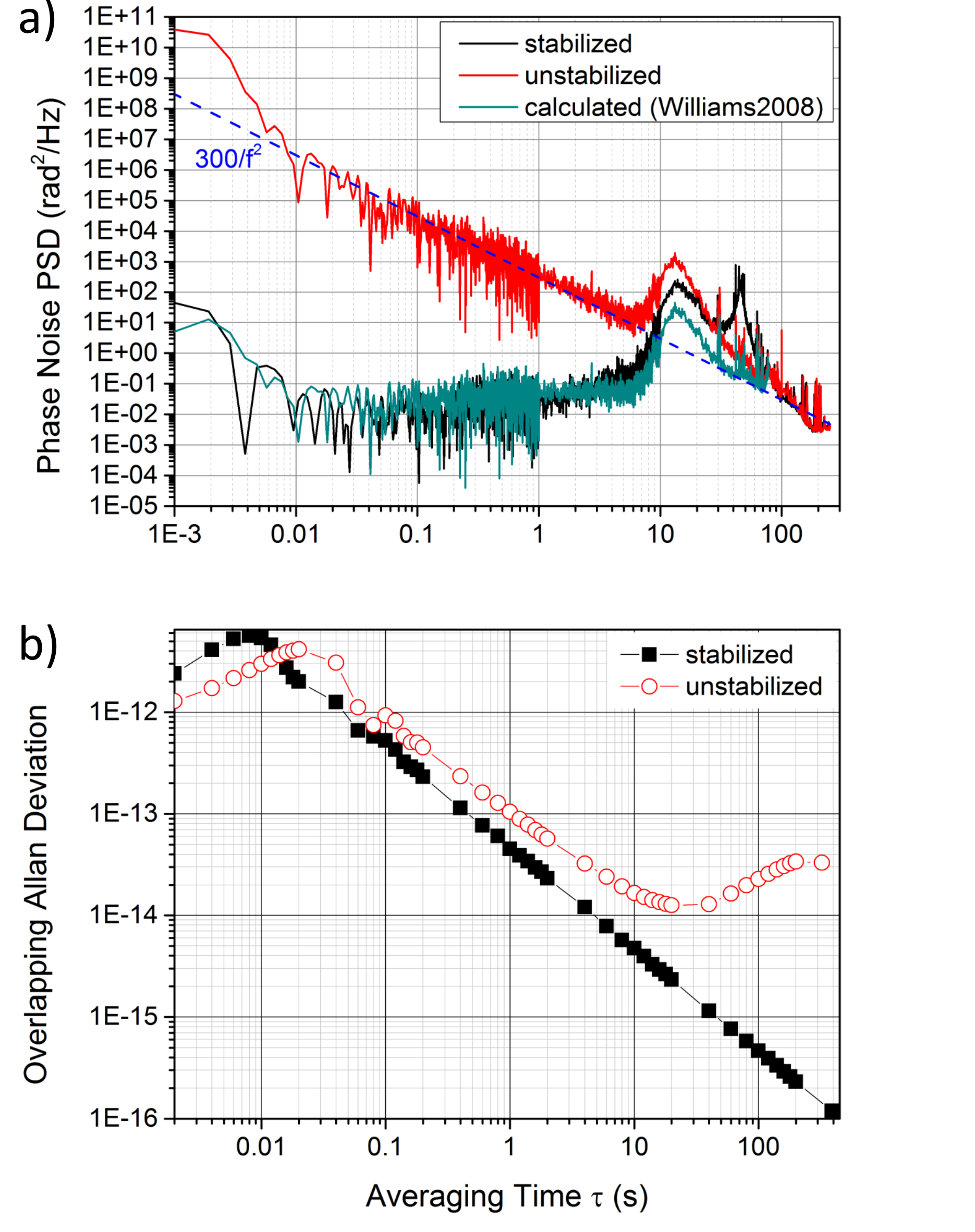}
\caption{The figure shows the phase noise (panel a) of the remote beat frequency for the loop being unstabilized and stabilized, respectively, as well as the corresponding short-term instabilities (Allan deviation, panel b). The data are obtained using a K\&K frequency counter in $\Pi$-mode. Also shown is the expected phasenoise for the stabilized case as calculated from the phase noise of the unstabilized loop according to \cite{williams2008}.} 
\label{fig:phasenoise}
\end{figure*}

\begin{figure*}[htbp]
\centering\includegraphics[width=8cm]{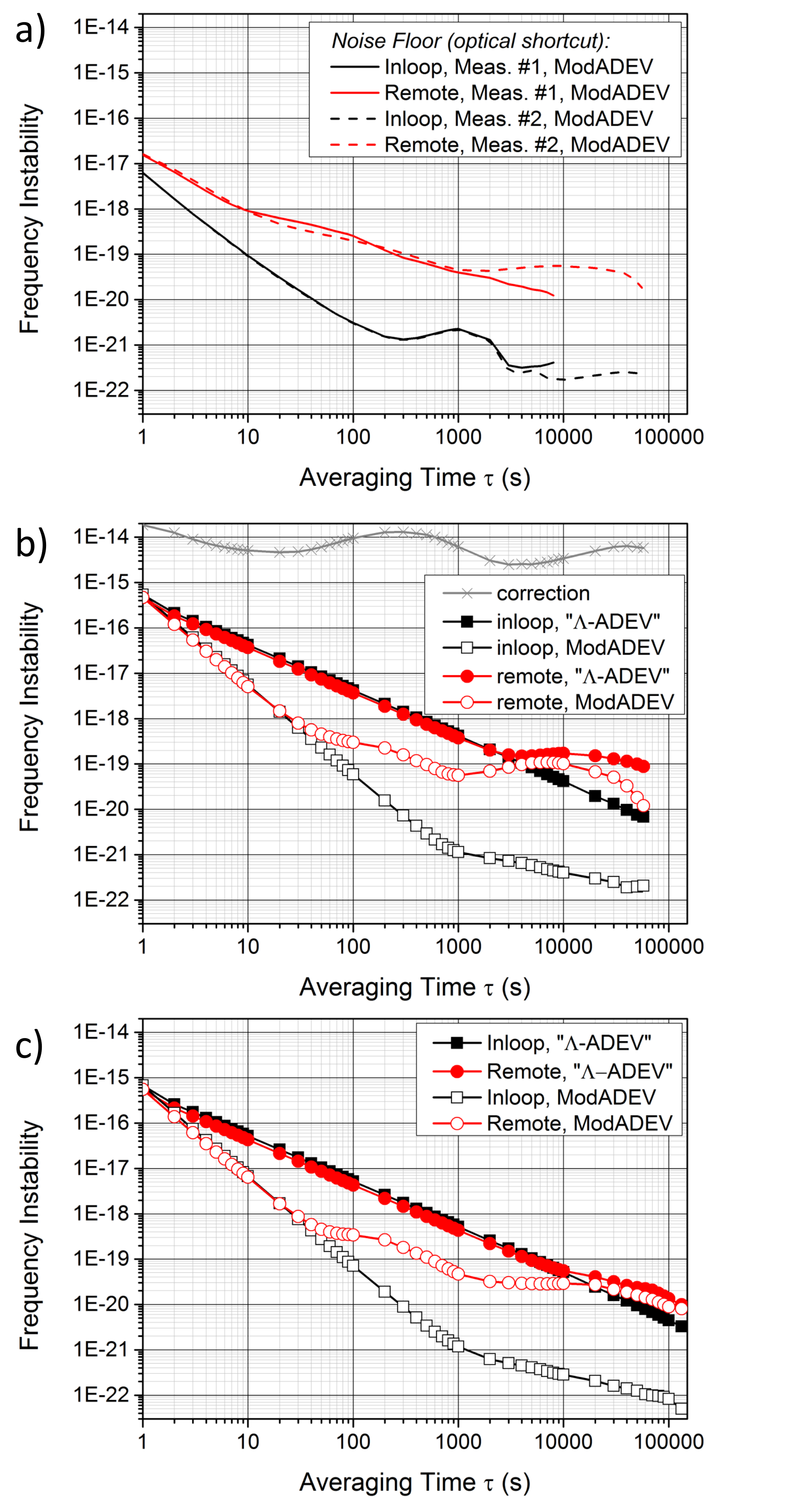}
\caption{Shown are the fractional frequency instabilities; all data are taken on a deadtime-free K\&K-counter operated in $\Lambda$-averaging mode with 1~s gate time. Panel a) shows two measurements of the noise floor (modified Allan deviation) for the loop being short-cut in the lab; panel b) and c) show the frequency instabilities for two measurements of the optical frequency transfer over the 660 km loop; the time between the measurements is several weeks. The measurements (modified Allan deviation) after about 100 s reach the noise floor at around $3\times10^{-19}$. Also shown in panel b) is the frequency instability of the unstabilized loop as obtained from the correction signal of the stabilization.} 
\label{fig:instability}
\end{figure*}

To characterize the optical transfer over the 660 km loop, we first measure the phase noise of the “remote” beatnote both for the case of active stabilization and for the case of a fixed frequency of acousto-optical modulator ``$\omega_1$''.  The results are shown in figure \ref{fig:phasenoise}, where the phase noise was obtained from the phase data taken with a K\&K counter in $\Pi$-mode at a gate time of 2 ms. The signal-to-noise ratio (100 kHz resolution bandwidth) is around 32 dB for the remote beat and around 27 dB for the roundtrip inloop beat.
\\Shown in fig. \ref{fig:phasenoise} is the noise of the unstabilized link. Between 0.1 Hz and about 10 Hz it falls off as 300/$f^2$, where $f$ is the Fourier frequency. At around 10 Hz to 15 Hz the phase noise exibits a broad maximum, which we attribute to infra-sound perturbations of the fiber, arising from road traffic \cite{sliwczynski2014}, or from air pipes in climate controls. Beyond this maximum the noise falls off more steeply.
\\In the stabilized case a servo bump of the link stabilization is visible at around 45 Hz, which is lower than expected from the path delay alone. This effect is still under investigation. The phase noise is suppressed within this bandwidth, achieving white phasenoise below 10 Hz, close to the calculated limit \cite{williams2008}. Figure \ref{fig:phasenoise}b) shows the corresponding plots of the frequency instability (Allan deviation). In the stabilized case, for averaging times longer than 0.1 s (10 Hz) it falls off as $a/\tau$ as is expected for white phase noise, with $a = 4.5\times10^{-14}$ and $\tau$ being the averaging time.
\\Figure \ref{fig:instability} shows measurements of the frequency instability for longer averaging times, this time taken with the counters operating in $\Lambda$-averaging mode with 1~s gate time.
\\The noise floor was measured after optically “short-cutting” the stabilized link in the lab, i.e. connecting the output of the local acousto-optic modulator to the input of the “remote” EDFA using optical attenuators, see fig. \ref{fig:instability}a). It is around $1.6\times10^{-17}$ at an averaging time of 1~s and reaches a floor in the $10^{-20}$-range for a few hundreds of seconds of averaging time. We attribute the variation in the $10^{-20}$ range to differential temperature fluctuations in the local and the ”remote” optical setup. Therefore, we regard observations of instabilities below $10^{-19}$ as fortuitous in the present setup, where the temperature of the optical setups is not controlled actively.
\\The measurements of the transfer instability via the 660~km loop, figs. \ref{fig:instability}b)+c), were taken over a  period of around 2.5 days, fig. \ref{fig:instability}b), and over a period of around six days, fig. \ref{fig:instability}c), respectively. The measurements are separated by about one month. No data points were removed within the measurement periods.
\\Figure \ref{fig:instability}b) shows (crosses) the instability of the free-running link as given by the correction signal of the link stabilization. It undulates in the $10^{-15}$ range with an overall mean of the fractional frequency offset of here $-1.7\times10^{-14}$. The instability curve indicates processes with periods of around 500 s and around one day, respectively. We attribute the slow perturbation to daily temperature variations along the link, while the shorter-period process might by due to cycles of the climate control e.g. at Kassel University.
\\The filled symbols represent the result of applying the Allan deviation formalism to the $\Lambda$-type data, corresponding to an unweighted averaging of the $\Lambda$-weighted 1~s values. The Allan deviations fall off as $5\times10^{-16}/\tau$ as expected e.g. for white phase noise. For the first measurement, a flicker floor around $1\times10^{-19}$ seems to be reached after about 3000~s of averaging time $\tau$; for the second measurement the Allan deviation averages down to around $10^{-20}$. Also shown is the modified Allan deviation (ModADEV) for the data (open symbols), corresponding to overlapping ``$\Lambda$''-type weighting. It falls off with $1/\tau^{3/2}$ for averaging times $\tau$ up to 20~s, confirming the dominance of white phase noise. At an averaging time of around 100~s it reaches an instability (ModADEV) of $3\times10^{-19}$, and of $1\times10^{-19}$ for an averaging time of around 400~s. The unweighted mean of the fractional frequency offset of the first measurement (fig. \ref{fig:instability}b) is $-2.1\times10^{-21}$ for the inloop beat and $1.1\times10^{-19}$ for the remote beat. For the second measurement (fig. \ref{fig:instability}c), the unweighted means are $1.1\times10^{-21}$ and $-2.2\times10^{-20}$ for the inloop and remote beat, respectively. In both cases, the mean of the remote beat frequency offset is slightly larger than the last point of the ``$\Lambda$''-Allan deviation for unweighted averaging. However, as stated above, this is to be expected as the instability (ModADEV) for averaging times beyond a few tens of seconds is limited by the current noise floor. The variable noise floor, which for averaging times beyond around 300~s is below $10^{-19}$, indicates out-of-loop noise uncorrelated between the local and the ``remote'' optical setup. 

\section{Summary}     
This paper presents for the first time results from long-haul ultrastable optical frequency transfer using a fiber Brillouin amplifier as the only means of remote amplification, installed along the link. The link is formed by a pair of fibers forming a 660 km loop, where the roundtrip inloop signal experiences a total of five Brillouin amplifications. The fibers are part of a new connection Braunschweig-Strasbourg-Paris. We obtained continuous measurement intervals of several days without observing cycle slips of the detected signal. 
\\We find that the phase noise we observe is slightly larger but comparable to the noise we have observed on a different, 920 km link between Braunschweig and Garching, including an ``infra-sound'' bump between 10 Hz and 20 Hz.  Between 10 mHz and 10 Hz the phase noise is observed to fall off as $300/f^2$~rad$^2$/Hz. For the stabilized case we find that the frequency of the servo bump is lower than would be expected from the calculations; the reason of this discrepancy is under investigation.   
\\For the transfer of an optical continuous-wave frequency we find an instability better than $10^{-19}$ and an accuracy of $1\times10^{-19}$. Below $10^{-19}$ we are currently limited by uncorrelated noise. We attribute this to uncorrelated temperature fluctuations leading to out-of-loop noise between the local and the “remote” setup, as presently they  do not have an active temperature control. 
\\The results demonstrate, that remote Brillouin amplification is well-suited for transferring an ultrastable, cw optical frequency. Compared to EDFA as the traditional means of amplification, it allows operating at larger gain and thus allows a larger distance between the amplifiers. Here, the maximum distance bridged between amplifications was around 250~km.
\\Currently a second remote Brillouin amplifier is being set up to investigate the performance of a link containing a chain of remote Brillouin amplifiers. This would then extend the length of the loop to around 1000~km.
\section*{Acknowledgements}
We are most grateful to Wolf-Christian K\"onig and Fritz Hack from GasLine GmbH, Karsten Leipold, Stefan Piger and Christian Grimm from Deutsches Forschungsnetz e.V., as well as Thomas Vetter, Harald Klatte und Ugur Koc from Kassel University for their invaluable support during planning and preparation of the measurements as well as the installation and maintenance of the remote Brillouin amplifer. We thank Andr\'e Uhde, J\"orn Falke, Matthias Misera, and Marion Wengler for excellent technical support.  This work is supported by the European Metrology Programme (EMRP) SIB-02 (“NEAT-FT”). The EMRP is jointly funded by the EMRP participating countries within EURAMET and the European Union.

\clearpage
\listoffigures
\clearpage

\end{document}